\documentclass[12pt]{article}
\usepackage{amsmath} 
\usepackage{longtable}
\usepackage{graphicx,epsfig}    
\usepackage{color}
\usepackage{cite}
\usepackage{amssymb}
\usepackage[utf8]{inputenc}
\usepackage{appendix}
\usepackage{array}
\newcommand{\as}{\left(\frac{\alpha_s}{4 \pi}\right)}
\newcommand{\aq}{\left(\frac{\alpha}{4 \pi}\right)}
\newcommand{\lren}{\log\left(\frac{\mu_R^2}{\mu_F^2}\right)}
\newcommand{\FacLog}{L_{\mu_F}\;}
\newcommand{\FacLogDos}{L^2_{\mu_F}\;}
\newcommand{\DiLog}{{\rm Li}_2}
\newcommand{\TriLog}{{\rm Li}_3}
\newcommand{\Spence}{{\rm S}_{1,2}}
\newcommand{\Dcero}{\mathcal{D}_0 (x)}
\newcommand{\Duno}{\mathcal{D}_1 (x)}
\newcommand{\Ddos}{\mathcal{D}_2 (x)}
\newcommand{\Dtres}{\mathcal{D}_3 (x)}

\begin{document}
\setlength{\parskip}{0.15cm}
\setlength{\baselineskip}{0.52cm}
\begin{titlepage}

\begin{flushright}
ICAS 033/18
\end{flushright}

\renewcommand{\thefootnote}{\fnsymbol{footnote}}
\thispagestyle{empty}
\noindent

\vspace{0.5cm}

\begin{center}
{\bf \Large 
QCD$\oplus$QED  NNLO corrections to Drell Yan production\\}
  \vspace{1.25cm}
{\large
Daniel de Florian\footnote{deflo@unsam.edu.ar},
Manuel Der\footnote{mder@unsam.edu.ar} and 
Ignacio Fabre\footnote{ifabre@unsam.edu.ar} \\
}
 \vspace{1.25cm}
 {
International Center for Advanced Studies (ICAS), ECyT-UNSAM,\\ 
Campus Miguelete, 25 de Mayo y Francia, (1650) Buenos Aires, Argentina \\
 }
  \vspace{1.5cm}
  \large {\bf Abstract}
  \vspace{-0.2cm}
\end{center}

We compute the  QCD$\times$QED (${\cal{O}}(\alpha_s \alpha)$) mixed and QED$^2$ (${\cal{O}}(\alpha^2)$)  corrections to the production of an on-shell $Z$ boson in hadronic collisions. We obtain them by profiting from the calculation of the pure QCD terms after taking the corresponding abelian limits. Therefore,  we extend the available knowledge up to complete next-to-next-to leading order precision in QCD$\oplus$QED. 

We present explicit results for the  perturbative coefficients and perform the phenomenological analysis at different collider energies with particular emphasis on the mixed corrections. We study the contribution from the different channels and discuss the scale dependence stabilisation effect. We consider a factorisation approximation for the mixed order terms  and show that it fails to reproduce the exact result. We find that the contributions are small, typically at the few per mille level, but that under some kinematical conditions they can compete with the pure QCD NNLO ones.

\hfill

\end{titlepage}
\setcounter{footnote}{1}
\renewcommand{\thefootnote}{\fnsymbol{footnote}}
%
\section{Introduction}
\label{sec:intro}

In recent years the development of high precision experiments in particle physics demanded a theoretical upgrade to match the accuracy achieved at the LHC. The state-of-the-art in fixed order computations for processes with up to two hard partons in the final state is reaching next-to-next-to-leading-order (NNLO), i.e. ${\cal{O}}(\alpha_s^2)$. Since $\alpha_s^2\sim \alpha$, it becomes necessary to include also the corresponding NLO ElectroWeak (EW) corrections, that for many observables exceed the few percent level (e.g. \cite{Dittmaier:2012kx,Dittmaier:2013hha}) and become quantitatively important for an accurate description.
Both  precise measurements and  calculations are essential to test different aspects of the Standard Model (SM) and to discern between them and possible new physics evidences due to Beyond the Standard Model (BSM)  effects.

In this sense, inclusive massive lepton pair production (Drell-Yan process) has worked as an important testground of perturbative quantum chromodynamics (QCD). On one hand, it has offered a sensitive way to study parton distribution functions (PDFs) \cite{Ball:2014uwa,Harland-Lang:2014zoa,Dulat:2015mca}. From weak bosons production, charge asymmetry measurements and invariant mass dependences have helped to extract precise information on both the quarks valence structure functions and the separation of quarks flavours, as well.
On the other hand, knowing the behaviour of charged-current (CC) and neutral-current (NC) processes has allowed to perform high-precision measurements of fundamental ElectroWeak 
parameters, like $Z$ and $W$ widths and masses and the EW mixing angle. With $W$ and $Z$ bosons in final or intermediate states of hadronic processes, some clean production channels can be characterised with great level of accuracy by studying leptonic decay modes. This has also served to accurately calibrate detector components, which was important for further measurements \cite{Brock:1999ep}.

In addition, the Drell-Yan process is not only relevant to test SM predictions, but also to evaluate alternative BSM theories, where $W$ and $Z$ bosons usually appear as final or intermediate states in the decay of particles predicted in new physics models, like new gauge interactions, supersymmetry or heavy resonances \cite{Mangano:2015ejw}. For all these reasons, the study of Drell-Yan processes, and in general, theoretical and experimental exploration of multiple EW gauge boson production, becomes a priority in the development of current high-energy physics.

In this sense, and considering that the improvement in statistics over the last years has made higher-order corrections experimentally noticeable, having access to QCD (and QED) corrections to these processes has become of great importance to  put the previous predictions on a firmer ground. For instance, it is important to achieve a better comprehension of the so-called Sudakov regime, where large logarithmic EW factors play an important role, and $\mathcal{O}(\alpha\alpha_s)$, that represent the first QCD corrections to these large EW factors, or even $\mathcal{O}(\alpha^2)$ corrections,  might be sizable. 

Furthermore, recent work has been performed to include QED effects in the evolution of parton distributions, by providing explicit expressions for splitting kernels up to  $\mathcal{O}(\alpha\alpha_s)$  \cite{deFlorian:2015ujt} and $\mathcal{O}(\alpha^2)$\cite{deFlorian:2016gvk} and by the determination of precise photon distributions in the proton within the LUXqed approach \cite{Manohar:2016nzj,Manohar:2017eqh} . The availability of these QED corrected parton distributions is essential to match the theoretical calculations at the partonic level.

From the point of view of partonic cross-sections, the $\mathcal{O}(\alpha\alpha_s)$ and $\mathcal{O}(\alpha)$  corrections represent the first EW and mixed order contributions to Drell-Yan pair production in the general expansion
\begin{equation}
\label{eq:expansion}
d\sigma = \sum _{i,j} \alpha_s^i \alpha^j d\sigma^{(i,j)},
\end{equation}
where pure EW $d\sigma^{(0,j)}$ and QCD $d\sigma^{(i,0)}$ corrections, as well as  \textit{mixed order} contributions, which combine effects of the two interactions, arise.

So far, QCD corrections to the total cross- section have been calculated at next-to leading order (NLO) in ref \cite{Altarelli:1979ub}, and at next-to-next-to leading order (NNLO) in an inclusive way, in refs. \cite{Hamberg:1990np,vanNeerven:1991gh,Harlander:2002wh}.  Exclusive results have also been presented up to NNLO QCD accuracy \cite{Catani:2009sm,Melnikov:2006di,Melnikov:2006kv, Gavin:2010az,Gavin:2012sy,Boughezal:2016wmq}. Additionally, threshold calculations have been performed at next-to-next-to-next-to-leading-order (${\rm N^3LO}$) and next-to-next-to-next-to-leading-logarithmic (${\rm N^3LL}$) accuracy in refs. \cite{Ahmed:2014cla,Catani:2014uta}. 

On the other hand, concerning the EW contributions, exclusive computations for NLO-EW corrections to CC-DY are available in refs. \cite{Dittmaier:2001ay,Baur:2004ig,CarloniCalame:2006zq} and for NC-DY, in refs.  \cite{Baur:2001ze,Baur:1997wa}. Finally, progress towards the computation of NNLO-EW has been accomplished in recent years too \cite{Actis:2006ra,Actis:2006rb,Actis:2006rc,Degrassi:2003rw}.
Due to the lack of the full calculation of the NNLO mixed-order terms $\mathcal{O}(\alpha\alpha_s)$, different approaches have been followed to approximately combine the QCD and QED/EW corrections \cite{Cao:2004yy,Balossini:2009sa,Adam:2008pc,Li:2012wna,Barze:2013fru}, by either assuming the full factorisation or the additive combination of the strong and electroweak contributions.  Particularly, recent partial exclusive results have been presented for the resonance region, by using the pole approximation \cite{Dittmaier:2014koa,Dittmaier:2014qza,Dittmaier:2015rxo}.

The contributions for a general (i.e. including the decay of the gauge boson) perturbative calculation 
of Drell-Yan can be roughly characterised into the following subsets: on one hand, {\it purely factorisable} 
terms that arise due to
initial state ({\it production}, from the initial state partons) and final state ({\it decay}, from the final state leptons) emission
and, on the other hand, {\it non-factorisable} terms
originated by soft photon exchange between the production and the decay. 
The non-factorisable $\mathcal{O}(\alpha \alpha_s)$ terms have been shown \cite{Dittmaier:2014qza,Dittmaier:2014koa,Dittmaier:2015rxo} 
to have a negligible impact on the cross section, allowing to treat effectively  Drell-Yan  
in the (resonant) limit of the decoupling between the production and decay processes, at least for the achieved experimental accuracy. The results presented in  \cite{Dittmaier:2015rxo} also rely on the assumption that the missing {\it initial-initial state factorisable } $\mathcal{O}(\alpha \alpha_s)$ contributions are very small.

The computation of the so far unknown mixed  QCD$\times$QED $\mathcal{O}(\alpha \alpha_s)$ corrections
to the inclusive on-shell production of a $Z$ boson in hadronic collisions 
 is exactly the main goal of this paper\footnote{In order to separate the QED contributions computed here from the {\it weak} induced effects, we consider the coupling between the $Z$ boson and the quarks as an {\it effective coupling} and do not take into account  self-energy insertions in the $Z$ (and eventually $\gamma$) propagator.}.
Those contributions are by themselves a gauge-invariant set of the complete Drell-Yan cross section calculation 
at  $\mathcal{O}(\alpha \alpha_s)$, even for CC-DY. Furthermore, counting with analytical expressions for the total cross section can be useful to establish a subtraction method to compute differential distributions for different observables at $\mathcal{O}(\alpha \alpha_s)$ by extending, for example, the $q_T-$ subtraction method \cite{Catani:2007vq} originally developed for pure QCD corrections.

In principle a full computation of QCD$\times$QED $\mathcal{O}(\alpha \alpha_s)$ terms involves, as in any NNLO calculation, the evaluation
of double-virtual, single-virtual plus one parton emission and double parton emission contributions, where parton
in general refers to  quarks, antiquarks, gluons, and photons. Most of the needed double real contributions where recently presented in \cite{Bonciani:2016wya}, including the case of $W$ boson production which is not discussed in this paper \footnote{Only the contribution from the interference of QCD and QED $qq\rightarrow qq$ diagrams is missing in \cite{Bonciani:2016wya}.},while the master integrals for the two-loop calculation were obtained in \cite{Bonciani:2016ypc}.

Instead of following the path of a dedicated calculation for each term, in this work we profit from the
available computation of NNLO pure QCD corrections $\mathcal{O}(\alpha_s^2)$ presented in \cite{Hamberg:1990np} and, by pointing out the abelian
component, we extract the corresponding QCD$\times$QED $\mathcal{O}(\alpha \alpha_s)$ contributions and asses the
phenomenological impact for the inclusive cross section at different hadronic energies.

Furthermore, by following the same procedure
we also present the QED$^2$ $\mathcal{O}(\alpha^2)$ corrections, completing, therefore, the set of NNLO contributions 
in QCD$\oplus$QED (i.e. all terms that correspond to $i+j=2$ in Eq.\eqref{eq:expansion}).

This paper is organised as follows:
In Section \ref{sec:abel} we present the method used to compute the QCD$\times$QED $\mathcal{O}(\alpha \alpha_s)$ and QED$^2$ $\mathcal{O}(\alpha^2)$ contributions from the pure QCD corrections $\mathcal{O}(\alpha_s^2)$. In Section \ref{sec:results} we present the results and study the detailed phenomenology of the corrections at different collider energies. Finally, in Section \ref{sec:conc} we present our conclusions. The explicit NNLO coefficients are presented in the Appendices.

\section{Abelianisation procedure}
\label{sec:abel}
In order to achieve the $\mathcal{O}(\alpha\alpha_s)$ contributions we profit from the availability of previous NNLO QCD calculations \cite{Hamberg:1990np}. In summary, we analyse the contributing  diagrams for each interaction and  take the corresponding abelian limit  from the existing NNLO QCD calculation \cite{deFlorian:2015ujt,deFlorian:2016gvk}.
To schematise the procedure used to accomplish the mixed order contribution, we describe the algorithm of {\it gluon-photon interchange} taking as an example the most relevant  $q\bar{q}$ channel.

First of all we analyse all the topologies inherent to each partonic subprocess in QCD and determine the corresponding colour factors. Then we replace a gluon by a photon in each diagram and profit from the similarity in QED and QCD fermion coupling factors to pin down the abelian parts of both calculations. Furthermore, given that the photon has no colour, non-abelian terms (i.e. all the terms arising from NNLO QCD diagrams that contain three or four gluon vertices) do not contribute to QED corrections and therefore can be thrown out. Thus, we recalculate the colour factors for the mixed topologies attained identifying changes and substitutions to be made in previous QCD results in order to obtain mixed order QCD$\times$QED correction terms.

\begin{figure}
\includegraphics[width=0.7\columnwidth]{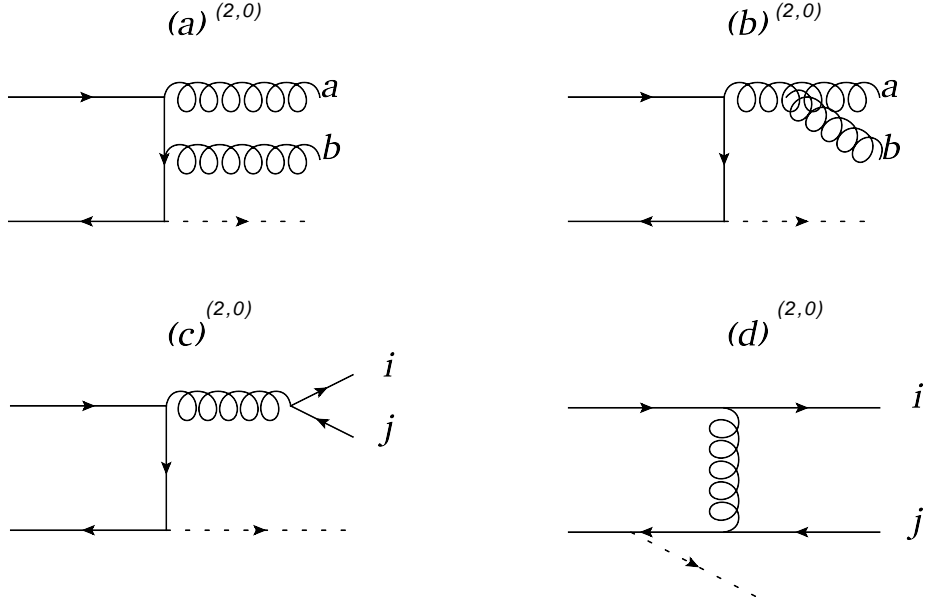}
\centering
\caption{Some of the diagrams contributing to NNLO QCD corrections to Drell-Yan.}
 \label{qqbar}
\end{figure}

For the explicit case of the  $q\bar{q}$ channel, in order to show how the method works and given that the colour factors in the partonic cross section only depend on the QCD structure, we concentrate on diagrams with double real emission which appear only to tree level (the same considerations can be applied to the two-loop and one-loop plus real emission contributions). As can be observed in Fig.\ref{qqbar} there are four kinds of diagrams to consider, which we will label with the supra-index $(k,l)$ according to the total number of QCD ($k$) and QED ($l$)  vertices for each topology. It is also important to note that the order of external momenta does affect the colour structure of the diagram. In this sense, we will refer as $(a)$ to the diagram showed in Fig.\ref{qqbar}, while   $(a')$ represents the corresponding diagram obtained after crossing the final state parton lines. There we can recognise different colour factors, according to the configurations of colour matrix traces in the calculation of each contribution to the partonic cross section.

For example,  terms corresponding to the calculation of $\left| (a)^{(2,0)} \right| ^2$ in NNLO QCD result proportional to $\frac{1}{2N_C^2}Tr\left[ T^bT^aT^aT^b\right]=\frac{1}{2N_C}C_F^2$, where the $N_C^2$ in the denominator arises due to the average over the colour factor of the incoming quarks and the symmetry factor $1/2$ is due to the appearance of two identical gluons in the final state.
For the case of $\left[(a)^{(2,0)} \times (a'^*)^{(2,0)}\right]$ both abelian and non-abelian contributions appear resulting in a factor $\frac{1}{2 N_C^2}Tr\left[ T^bT^aT^bT^a\right]=\frac{1}{2N_C}C_F\left(C_F-C_A/2\right)$ and, when considering terms from $\left[(b)^{(2,0)} \times (a^*)^{(2,0)}\right]$, they result proportional to  
$\frac{1}{2N_C^2}\ f^{abc}Tr\left[ T^cT^aT^b\right]=\frac{1}{2N_C^2}\ Tr\left[[ T^a,T^b]T^aT^b\right]=-\frac{1}{2N_C}(C_FC_A/2)$, a purely non-abelian contribution.  

\begin{figure}
\includegraphics[width=0.7\columnwidth]{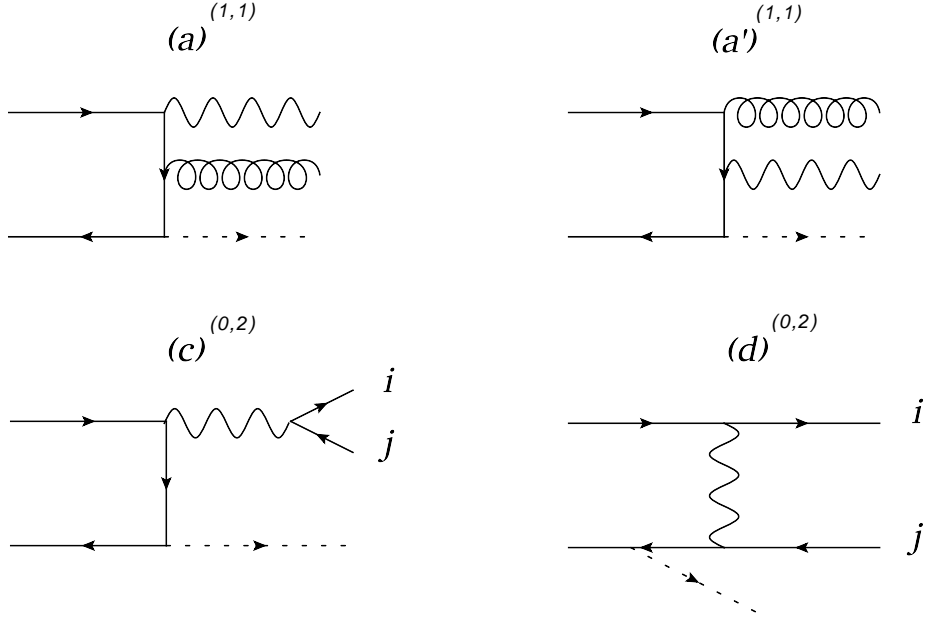}
\centering
\caption{Diagrams that result after applying the abelianisation procedure to the real NNLO QCD corrections  in Fig.\ref{qqbar}.}
 \label{qqbarQCDQED}
\end{figure}

Once colour factors are characterised for each term, we choose a gluon in the diagram, replace it by a photon  and recalculate the colour structure, thus obtaining modified diagrams with the corresponding new factors for QCD$\times$QED corrections. These are shown in Fig.\ref{qqbarQCDQED}. Naturally, all the diagrams of type $(b)$ (i.e. topologies containing at least one 3-gluon-vertex), which always contribute to second order for the NNLO QCD calculation in this process, vanish when considering the abelian limit.

Taking this into account, we find that the modified factors for $\left|(a)^{(1,1)}\right| ^2$  and $\left[(a)^{(1,1)} \times (a'^*)^{(1,1)}\right]$ are both given by  $\frac{e_q^2}{N_C^2}Tr[ T^aT^a]=\frac{e_q^2}{N_C} C_F$, where we have included the charge of the quark for the QED coupling,  while the non-abelian one  obviously vanishes. Here we may notice that all the colour factors proportional to $C_A$, which corresponds to non-abelian part of the calculation, could be thrown out when considering the abelian limit, while the ones proportional to $C_F^2$ are to be replaced by $2e_q^2C_F$, thus obtaining QCD$\times$QED factors in each case.
It is worth noticing by performing the same analysis for the topology shown in Fig.\ref{qqbar}c, i.e. the production of a $q\bar{q}$ pair, that also the colour factor $T_R$ vanishes when the similar contribution is analysed in the 
QCD$\times$QED case, since the result for $\left[(c)^{(2,0)} \times (c^*)^{(0,2)}\right]$ becomes proportional to $Tr[ T^a]$. Therefore, since terms proportional to both $C_A$ and $T_R$ are vanishing, the same occurs for terms proportional to $\beta_0^{QCD}$ in the original pure QCD calculation, consistent with the fact that no renormalisation is needed at this order either for the QED or QCD couplings \footnote{As stated above, we consider the Born coupling between the quarks and the $Z$ in the sense of an effective coupling.}.
Same wise, only a few contributions survive in the products of the type $\left[(c)^{(2,0)} \times (d^*)^{(0,2)}\right]$ and $\left[(d)^{(2,0)} \times (d'^*)^{(0,2)}\right]$, i.e. the interference of amplitudes with one photon and with one gluon exchange.

\begin{table}
\begin{center}
\begin{tabular}{|l| c| c| c|}
 \hline
 \multicolumn{4}{|c|}
 {Colour factors in $q\bar{q}$} 
 \\[1ex] 
 \hline
 diagram &$\alpha_s^2$&$\alpha\times\alpha_s$&$\alpha^2$\\ \hline & & &\\  
$\left|(a)\right|^2$  & $C_F^2$    &$2e_q^2C_F$&$e_q^4$\\[1ex] 
$ (d)\times (d^*)$  & $C_F T_R$  & $0$ & $C_A\ e_i^2e_j^2$ \\[1ex]  
$(c)\times (c^*)$  & $n_F\ C_FT_R$ &  $0$  & $ e_q^2 \left[ N_C\displaystyle\sum_{k\epsilon Q} e_k^2 +\displaystyle\sum_{k\epsilon L}e_k^2 \right]$\\[1ex] 
$(a)\times (a'^{*})$ & $C_F^2-\frac{C_F\ C_A}{2}$  &$2e_q^2C_F$&$e_q^4$\\[1ex] 
$ (d)\times (d'^*)$  & $C_F^2-\frac{C_F\ C_A}{2}$  & $2e_q^2C_F$ & $e_q^4$ \\[1ex] 
$ (b) \times (a^*)$  & $-\frac{C_F\ C_A}{2}$  &  $0$  & $0$ \\[1ex] 
$ (c)\times (d^*)$  & $C_F^2-\frac{C_F\ C_A}{2}$  &  $2e_q^2C_F$  & $e_q^4$ \\[1ex] 
\hline
\end{tabular}
\end{center}
\label{comparacqqbar}
 \caption{Colour factors corresponding to $q\bar{q}$ channel for each contribution to NNLO QCD$\oplus$QED corrections to Drell-Yan, up to an overall $\frac{1}{2N_C}$ factor. Focusing on $\alpha^2$ factors, the third column includes sums over sets of quark ($Q$) and lepton ($L$) final state charges, while $e_i$ and $e_j$ refer to different quark flavour charges in the scattering.}
\end{table}

This  strategy can be  extended for all the topologies  in $q\bar{q}$.
In Table \ref{comparacqqbar} we show the different colour factors (after factorising an overall factor of $1/2N_C$) for diagrams contributing to $\sigma^{(2,0)}$, and the resulting ones after the abelianisation procedure corresponding to $\sigma^{(1,1)}$. The  replacements in the colour structures needed to go from the NNLO QCD coefficients to  the QCD$\times$QED ones can be directly read from the entries in Table \ref{comparacqqbar}.

As an important feature, this method shows to be versatile in order to obtain NNLO QED corrections to Drell-Yan as well (i.e. the calculation of $\sigma^{(0,2)}$), if  a deeper abelian limit is considered in this case. Here, by turning two gluons into photons from the topologies of NNLO QCD calculation  one can recover correction terms up to second order in $\alpha$, thus completing the set of QCD$\oplus$QED NNLO corrections to Drell-Yan, in the sense of  Eq.(\ref{eq:expansion}). The corresponding colour factors (including electric charges of both quarks and leptons that might appear in the final state) 
 are also shown in Table \ref{comparacqqbar} for the $q\bar{q}$ channel.

The same occurs for other channels, after treating carefully the initial flux factor, which depends on the colour properties of initial state particles.  For instance, both $q\gamma$ and $qg$ contributions to $\sigma^{(1,1)}$ can be obtained from the $qg$ calculation for NNLO QCD corrections, by choosing the initial or final state gluon, respectively, to perform the abelianisation and following the procedure detailed above. 
Particularly, in the case of $\gamma g$ channel, we have performed the explicit calculation of the fixed order corrections, finding perfect agreement with the result obtained by applying the abelianisation procedure.

\section{Results and Phenomenology}
\label{sec:results}

In general the cross section can be written as 
\begin{equation}
\frac{d\sigma^Z}{dQ^2}=\tau \sigma_Z (Q^2,M_Z^2)W_Z(\tau , Q^2),  
\label{eq:Xsec}
\end{equation}
where $\sigma_Z$ is the point-like LO cross section, $\sqrt{S}$  is the hadronic  centre-of-mass energy, $Q$ the invariant mass of the produced $Z$, $\tau=\frac{Q^2}{S}$ and $W_Z(\tau, Q^2)$ is the hadronic structure function.

The point-like cross section that appears in Eq.\eqref{eq:Xsec} is defined as
\begin{equation}
\label{ec:sigmaLO}
\sigma_Z(Q^2, M_Z^2) = \frac{\pi \alpha}{4 M_Z \sin^2\theta_W \cos^2\theta_W} \frac{1}{N_C} \frac{\Gamma_{Z\rightarrow X}}{(Q^2-M_Z^2)^2 + M_Z^2 \Gamma_Z^2},
\end{equation}
where $N_C = 3$ is the number of quark colours, $\theta_W$ is the weak mixing angle (with $\sin^2 \theta_W=0.23$), $M_Z=91.187$ GeV and $\Gamma_Z$ are the mass and width of the $Z$, and $\Gamma_{Z\rightarrow X}$ is the partial width due to the decay of the $Z$ to $X$ (e.g. for leptonic decay, $X = \ell \bar \ell$). The narrow-width approximation used along this paper consists on making the following replacement
\begin{equation}
\frac{1}{(Q^2-M_Z^2)^2 + M_Z^2 \Gamma_Z^2}\rightarrow \frac{\pi}{M_Z \Gamma_Z} \delta(Q^2-M_Z^2).
\end{equation}
ensuring the decoupling of the production and decay mechanisms.
The hadronic structure function appearing in \eqref{eq:Xsec} can be written as a sum of contributions of different orders
\begin{equation}
\label{ec:w}
W_Z(\tau, Q^2) = \int_0^1 {\rm d}x_1\;\int_0^1 {\rm d}x_2\;\int_0^1 {\rm d}x\; \delta(\tau-x x_1 x_2)  \sum_{i,\,j} \as^i \aq^j \; w_Z^{(i,j)} (x, x_1, x_2, Q^2),
\end{equation}
where the dependence on the factorisation $\mu_F$ and renormalisation $\mu_R$ scales is implicit.

The analytic expressions for the inclusive cross section of Drell-Yan $Z$-production at QCD$\oplus$QED NNLO are presented in the Appendices. In this section we study the phenomenology of the total inclusive cross section, i.e. in all the decay channels of the $Z$, within the narrow-width approximation. To this end, a specific code was written which makes use of the LHAPDF\cite{Buckley:2014ana} package to interpolate 
sets of parton distribution functions.

\begin{figure}
\includegraphics[width=0.6\columnwidth]{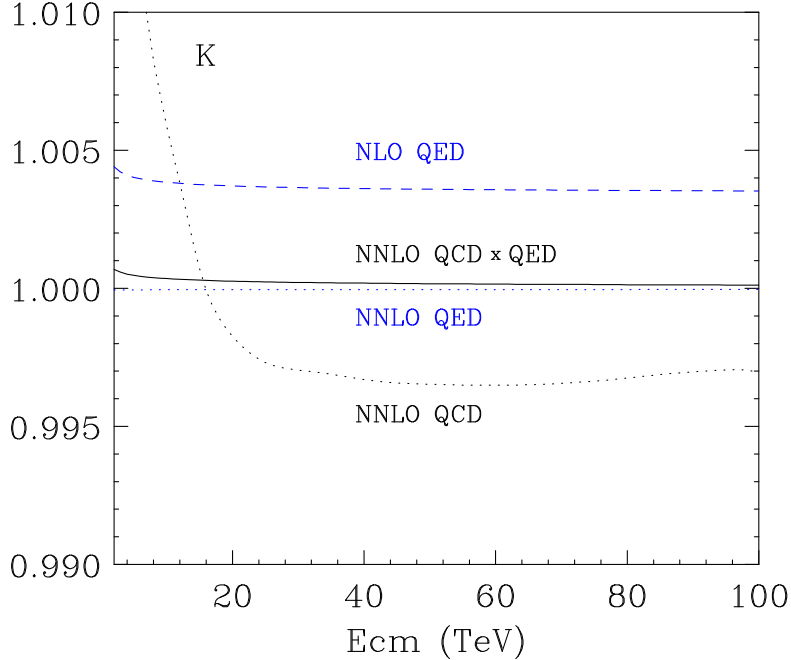}
\centering
\caption{$K$-factors for the different distributions as defined in Eq.(\ref{k-eq}). The (blue) dashed line corresponds to $K_{QED}^{NLO}$, the (blue) dotted line to  $K_{QED}^{NNLO}$, the solid line to the mixed $K_{QCD\times QED}^{NNLO}$ and the (black) dotted line to the pure NNLO QCD corrections  $K_{QCD}^{NNLO}$.}
 \label{kfactor}
\end{figure}

 For the phenomenological study, unless explicitly stated, we set the renormalisation and factorisation scales to $\mu_R=\mu_F=M_Z$. For both interactions, we set the running coupling at the corresponding renormalisation scale (i.e. $\alpha(M_Z)\sim \frac{1}{128}$ \footnote{For the sake of simplicity we make the same choice for the value of the coupling between quarks and the $Z$ boson in the Born cross section in Eq.(\ref{ec:sigmaLO}).}) and always use the parton distributions to NNLO (QCD) accuracy \cite{Butterworth:2015oua,Ball:2014uwa,Harland-Lang:2014zoa,Dulat:2015mca} with the corresponding QED corrections from LUXqed  \cite{Manohar:2016nzj,Manohar:2017eqh}.
 In Fig.\ref{kfactor}  we plot the $K$-factors for different orders as a way to quantify the size of the 
 QED and QCD corrections to Drell-Yan at different centre-of-mass energies.

 Here the $K$-factor 
 is defined as the ratio of the cross-section computed at a given order over the previous one, i.e.
 \begin{eqnarray}
 \label{k-eq}
 K_{QED}^{NLO}&=& \frac{\sigma^{(0,0)} + \alpha \, \sigma^{(0,1)}}{\sigma^{(0,0)}} \nonumber \\
 K_{QCD}^{NNLO}&=& \frac{\sigma^{(0,0)} + \alpha_s \, \sigma^{(1,0)} + \alpha_s^2 \, \sigma^{(2,0)}  }{\sigma^{(0,0)}+ \alpha_s \, \sigma^{(1,0)}} \\
 K_{QED}^{NNLO}&=& \frac{\sigma^{(0,0)} + \alpha \, \sigma^{(0,1)} + \alpha^2 \, \sigma^{(0,2)}  }{\sigma^{(0,0)}+ \alpha \, \sigma^{(0,1)}} \nonumber \\
 K_{QCD\times QED}^{NNLO}&=& \frac{\sigma^{(0,0)} + \alpha \, \sigma^{(0,1)}+ \alpha_s \, \sigma^{(1,0)} + \alpha \alpha_s \, \sigma^{(1,1)}  }{\sigma^{(0,0)}+ \alpha \, \sigma^{(0,1)}+ \alpha_s \, \sigma^{(1,0)}}. \nonumber 
\end{eqnarray}

\begin{figure}
\includegraphics[width=0.6\columnwidth]{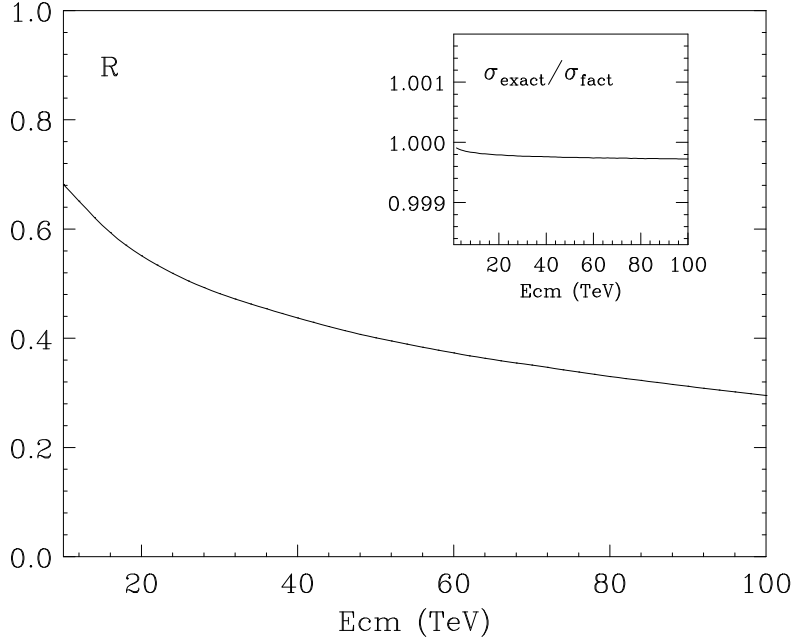}
\centering
\caption{Ratio $R$ between the exact and the factorisation approximation for the mixed QCD$\times$QED contributions. The inset plot shows the ratio of the cross section computed exactly and with the factorisation approximation for the mixed term.}
 \label{facto}
\end{figure}

 As can be observed, the NNLO QCD corrections are of the same ($\sim$ 5 per mille level) order, but typically with the opposite sign, as the NLO QED corrections, as expected from the simple counting $\alpha_s^2\sim \alpha$. The mixed QCD$\times$QED turn out to be positive and below the per mille level over the whole range of energies spanned in the plot. Interestingly, due to the particular dependence of the NNLO QCD corrections with the energy, with a sign change around $\sqrt{S}\sim 18$ TeV, for the LHC at 
 $\sqrt{S}\sim 14$ TeV the mixed QCD$\times$QED corrections are only a factor of $\sim 3.5$ smaller than the pure NNLO QCD contributions. Furthermore, for lower centre-of-mass energies $\sqrt{S}\sim 2$ TeV
 the mixed terms almost reach the per mille level and are just a factor of 5 smaller than the NLO QED ones, showing that the elementary counting of couplings can fail under certain kinematical conditions. The pure NNLO QED terms, also  plotted in Fig.\ref{kfactor}, are negative but the corrections always remain at the $\mathcal{O}(10^{-5})$ level.

Even though for this particular observable the mixed QCD$\times$QED contributions are small, it is interesting to study how well they can be approximated by the {\it factorisation} assumption on QED plus QCD corrections, where it is assumed that $\kappa_{\rm fact} = \left[K_{QED}^{NLO}\times K_{QCD}^{NLO}\right]_{\mathcal{O}(\alpha \alpha_s)} = \alpha\alpha_s\frac{\sigma^{(0,1)}  \sigma^{(1,0)}}{\sigma^{(0,0)}\sigma^{(0,0)}} $, compared to the exact case $\kappa_{\rm mixed} =\alpha\alpha_s\frac{\sigma^{(1,1)}}{\sigma^{(0,0)}}$.  For that purpose, in Fig.\ref{facto} we plot the following quantity
\begin{eqnarray}
 R= \frac{\kappa_{\rm mixed}}{\kappa_{\rm fact}} =\frac{\sigma^{(0,0)} \sigma^{(1,1)} } {\sigma^{(0,1)}  \sigma^{(1,0)}}, 
\end{eqnarray}
which is the ratio between the exact and the approximated factorised contribution.
As it can be observed, the factorisation approach fails to reproduce the correct behaviour of the mixed contribution typically by a factor of two or more. Of course, given the size of the corrections, the effect of the factorised treatment of these contributions is small at the level of the cross section, as shown in the inset plot of Fig.\ref{facto}, where we show the ratio between the cross section computed exactly and within the factorisation approach, but the situation might not hold for other observables or even for more exclusive distributions in Drell-Yan.

\begin{figure}[ht]
\includegraphics[width=0.6\columnwidth]{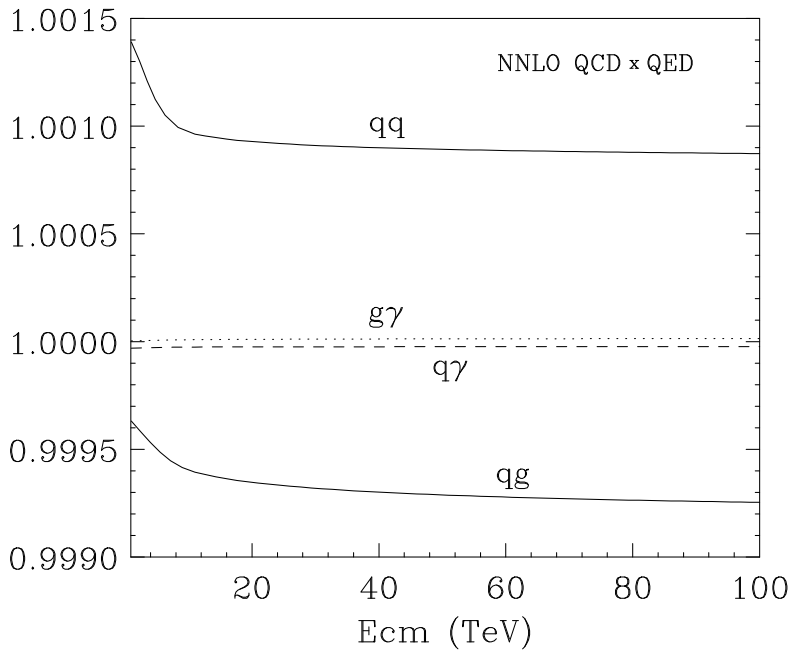}
\centering
\caption{Contribution to the mixed QCD$\times$QED $K$-factor from the different channels. Here the label $q$ accounts for both quarks and antiquarks and $qq$ represents the sum of $q\bar{q}$ and $qq$.}
 \label{canales}
\end{figure}
In Fig.\ref{canales} we show the contribution to the mixed QCD$\times$QED $K$-factor from the different channels. It is noticeable that the photon initiated contributions are rather small, mostly due to the size of the photon pdf in the proton, as can be observed by comparing $q\gamma$ and $qg$ contributions, which share the same partonic coefficient apart from the colour factor.
It is also clear that the different signs of $qq$ (fully dominated by the born level $q\bar{q}$ channel and exceeding the per mille level) and $qg$ contributions conspire to reduce the effect of the mixed QCD$\times$QED corrections to the Drell-Yan cross section. Again, in more exclusive distributions this partial cancellation might be spoiled by some kinematical cuts, resulting in an increase of the 
mixed order corrections.

\begin{figure}[th]
\includegraphics[width=0.6\columnwidth]{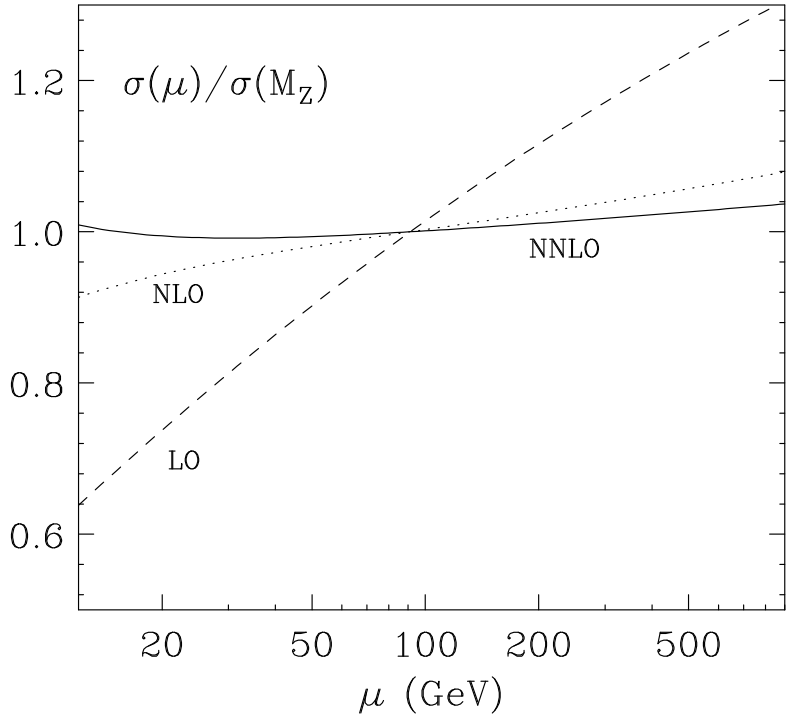}
\centering
\caption{Cross sections corresponding to LO (dashes, $i+j$=0 in Eq.(\ref{eq:expansion})), NLO(dots, $i+j$=0,1) and NNLO (solid, $i+j$=0,1,2) at different factorisation and renormalisation scales with $\mu_R=\mu_F=\mu$. All results are normalised by the corresponding cross section at $\mu=M_Z$.}
 \label{scales}
\end{figure}

Finally, we discuss the effect of the higher order contributions in the stabilisation of the perturbative expansion in terms of the scale dependence for $\sqrt{S}=13$ TeV (very similar behaviours are observed for other values of $\sqrt{S}$). In Fig.\ref{scales} we show the LO ($\sigma^{(0,0)}$), NLO ($\sigma^{(0,0)}+ \alpha \, \sigma^{(0,1)}+ \alpha_s \, \sigma^{(1,0)}$) and NNLO ($\sigma^{(0,0)} + \alpha \, \sigma^{(0,1)}+ \alpha_s \, \sigma^{(1,0)} + \alpha \alpha_s \, \sigma^{(1,1)} + \alpha^2 \, \sigma^{(0,2)}+ \alpha_s^2 \, \sigma^{(2,0)}$) cross sections for different values of the factorisation and renormalisation scales $\mu_R=\mu_F=\mu$, normalised by the corresponding value at the central scale $\mu=M_Z$. From the slope of the different curves, it is clearly visible the reduction in the scale dependence when including higher order corrections, mostly due to the dominant QCD effects but also thanks to the inclusion of the QED and mixed contributions.

\section{Conclusions}
\label{sec:conc}
In this article, mixed QCD$\times$QED as well as pure QED$^2$ NNLO corrections to the total Drell-Yan $Z$-production cross section were presented for the first time. This was achieved via an abelianisation procedure that profits from the available pure QCD NNLO result and proved to be a versatile technique.

We performed the phenomenological analysis finding that
the mixed corrections are  of the order of per mille at the LHC, but only a factor of $\sim 3.5$ smaller than the pure QCD NNLO due to a sign change that occurs in the latter at $\sqrt{S}\sim 14$ TeV. Pure QED NNLO terms are shown  to be negative corrections of the order of $10^{-5}$. The full QCD$\oplus$QED NNLO corrections are found to further stabilise the scale dependence of the final result.

The exact $K$-factor at order $\mathcal{O}(\alpha \alpha_s)$ was compared with the naive factorisation approximation, which consists of the mixed order term of the product of QCD and QED NLO $K$-factors. It was shown that the latter fails to reproduce the exact result by a factor of two or more. Although in this case the difference is not significant, due to the smallness of the overall contribution, this result hints that the factorisation approximation may not work for other processes nor for more exclusive measurements of the one presented herein.

\section*{Acknowledgments}

The work of D.deF. has been partially supported by Conicet, ANPCyT and the von Humboldt Foundation. We thank Massimiliano Grazzini, Leandro Cieri, German Sborlini and Javier Mazzitelli for discussions and comments on the manuscript.

\bibliography{biblio}

\newpage
\appendixtitleon
\appendixtitletocon
\begin{appendices}

\allowdisplaybreaks

\section*{Appendix A: Definitions and Notation}
\renewcommand{\theequation}{{\rm{A}}.\arabic{equation}}
\setcounter{equation}{0}

In this appendix we present for the first time, the coefficients $w^{(1,1)}$ and $w^{(0,2)}$ in the sense of Eq.(5). They include the NNLO mixed QED$\times$QCD and QED corrections, respectively.

\begin{align}
w_Z^{(1,1)} &= \sum_{i \in Q, \bar Q} q_i(x_1) \bar q_i(x_2) c_i 2 e_i^2 C_F \Delta_{q\bar q}^{(2) C_F}(x) \nonumber\\
& + \sum_{i \in Q, \bar Q} q_i(x_1)  q_i(x_2) c_i 2 e_i^2 C_F \Delta_{q q}^{(2) \text{id}}(x) \nonumber\\
& + \sum_{i \in Q, \bar Q}\left[2 C_A C_F (q_i(x_1) \gamma(x_2) + \gamma(x_1) q_i(x_2))+(q_i(x_1) g(x_2)+ g(x_1) q_i(x_2)) \right]\nonumber\\
&\qquad\quad\times c_i e_i^2 \Delta_{q g}^{(2) C_F}(x)  \nonumber\\
& + \left(g(x_1) \gamma(x_2)+\gamma(x_1) g(x_2)\right) 2 C_A \left(\sum_{k \in Q} c_k e_k^2  \right)\Delta_{g g}^{(2)}(x)\\
w_Z^{(0,2)} &= \sum_{i \in Q, \bar Q} q_i(x_1) \bar q_i(x_2)  \nonumber\\
& \times \Bigg\lbrace c_i  e_i^2 \left(e_i^2 \Delta_{q\bar q}^{(2) C_F}(x) +  2 \left[N_C \sum_{k \in Q} e_k^2 + \sum_{k \in L} e_k^2\right] \Delta_{q\bar q}^{(2) n_F}(x) + \beta_0^{QED} \Delta_{q\bar q}^{(1)}(x) \lren\right) \nonumber\\
& + \sum_{k \in Q} \left[ c_k \Delta_{q\bar q}^{(2) f}(x) + a_i a_k \Delta_{q\bar q}^{(2) ax}(x) \right]  2 C_A e_i^2 e_k^2\nonumber\\
& + \sum_{k \in L} \left[ c_k \Delta_{q\bar q}^{(2) f}(x) + a_i a_k \Delta_{q\bar q}^{(2) ax}(x) \right]  2 e_i^2 e_k^2 \Bigg\rbrace\nonumber\\
& + \sum_{i \in Q, \bar Q} c_i q_i(x_1)  q_i(x_2) \Delta_{q q}^{(2) \text{id}}(x) e_i^4\nonumber\\
& + \sum_{i,j \in Q, \bar Q} q_i(x_1) q_j(x_2) 2 C_A e_i^2 e_j^2 \nonumber\\
&\qquad\qquad\times\left[ (c_i + c_j) \Delta_{q q}^{(2)\text{non-id}}(x) + v_i v_j \Delta_{q_i q_j}^{(2)\text{non-id, V}}(x) +a_i a_j \Delta_{q q}^{(2)\text{non-id, A}}(x)\right]\nonumber\\
& + \sum_{i \in Q, \bar Q} (q_i(x_1) \gamma(x_2)+ \gamma(x_1) q_i(x_2)) c_i 2 C_A e_i^2 \left[ e_i^2 \Delta_{q g}^{(2) C_F}(x) + \beta_0^{QED} \Delta_{q g}^{(1)\gamma}(x) \lren\right]\nonumber\\
& + \gamma(x_1) \gamma(x_2)4 C_A \left[N_C \sum_{k \in Q} c_k e_k^4 + \sum_{k \in L} c_k e_k^4\right]\Delta_{g g}^{(2)C_F}(x) ,
\end{align}
where $Q\,(\bar Q)$ and $L$ are the sets of (anti)quarks and leptons considered (e.g. $Q = \{u,d,c,s,b\}$, $L=\{e,\mu,\tau\}$), $e_k$ is the charge of the particle $k$ in units of the electron charge, $\mu_F$ and $\mu_R$ are the factorisation and renormalisation scales,  $c_i = v_i^2 + a_i^2$ with $v_i$ and $a_i$ defined as the vector and axial couplings of particle $i$:
\begin{align}
v_u =& 1-\frac{8}{3} \sin^2\theta_W, &a_u =& -1,\\
v_d =& -1+\frac{4}{3} \sin^2\theta_W, &a_d =& 1,\\
v_e =& -1+4 \sin^2\theta_W, &a_e =& 1,
\end{align}
and replicated through families, the QED and QCD beta functions are given by
\begin{align}
\beta_0^{QED} =-\frac{2}{3}(N_C \sum_{k\in Q} e_k^2 + \sum_{k\in L} e_k^2)& &\beta_0^{QCD}=\frac{11}{3} C_A - \frac{2}{3} n_F,
\end{align}
$n_F$ is the number of quark flavours considered ($n_F = \#Q$), and the various $\Delta(x)$ corrections functions are defined in Appendix B.

The other coefficients needed for the full NNLO calculation, $w^{(i,0)}$ for $i\le2$ and $w^{(0,1)}$, were presented in \cite{Hamberg:1990np} and \cite{Baur:1997wa,Baur:2001ze} respectively. Here they are rewritten in this notation for the sake completeness and as a reference for the full set of NNLO corrections to the Drell-Yan $Z$ production.
\begin{align}
 w_Z^{(0,0)} =&  \sum_{i \in Q, \bar Q} c_i q_i(x_1) \bar q_i(x_2)  \delta(1-x)\\
w_Z^{(1,0)} =& \sum_{i \in Q, \bar Q}\big\lbrace c_i q_i(x_1) \bar q_i(x_2) C_F \Delta_{q\bar q}^{(1)} (x)\nonumber\\
& \left(q_i(x_1) g(x_2) + g(x_1) q_i(x_2)\right) c_i \Delta_{q g}^{(1)}(x)   \big \rbrace\\
w_Z^{(0,1)} =& \sum_{i \in Q, \bar Q} \big \lbrace c_i q_i(x_1) \bar q_i(x_2) e_i^2 \Delta_{q\bar q}^{(1)} (x)\nonumber\\
& \left(q_i(x_1) \gamma(x_2) + \gamma(x_1) q_i(x_2)\right) c_i 2 C_A e_i^2 \Delta_{q g}^{(1)}(x)   \big \rbrace\\
w_Z^{(2,0)} =& \sum_{i \in Q, \bar Q} q_i(x_1) \bar q_i(x_2) \nonumber\\
&\times \Bigg \lbrace  c_i  C_F \left(C_A  \Delta_{q\bar q}^{(2) C_A}(x) + C_F \Delta_{q\bar q}^{(2) C_F}(x) +  n_F \Delta_{q\bar q}^{(2) n_F}(x) + \beta_0^{QCD} \Delta_{q\bar q}^{(1)}(x) \lren\right) \nonumber\\
& + \sum_{k \in Q} \left[ c_k \Delta_{q\bar q}^{(2) f}(x) + a_i a_k \Delta_{q\bar q}^{(2) ax}(x) \right]  C_F \Bigg \rbrace\nonumber\\
& + \sum_{i \in Q, \bar Q} c_i q_i(x_1)  q_i(x_2) \Delta_{q q}^{(2) \text{id}}(x) C_F \left(C_F - \frac{1}{2} C_A\right)\nonumber\\
& + \sum_{i,j \in Q, \bar Q} q_i(x_1) q_j(x_2) C_F \left[ (c_i + c_j) \Delta_{q q}^{\text{(2) non-id}}(x) + v_i v_j \Delta_{q_i q_j}^{(2) \text{non-id, V}}(x) +a_i a_j \Delta_{q q}^{(2) \text{non-id, A}}(x)\right]\nonumber\\
& + \sum_{i \in Q, \bar Q} (q_i(x_1) g(x_2)+ g(x_1) q_i(x_2)) c_i\nonumber \\ 
&\qquad\qquad\qquad \times\left[C_A \Delta_{q g}^{(2) C_A}(x)+C_F \Delta_{q g}^{(2)C_F}(x) + \beta_0^{QCD} \Delta_{q g}^{(1)}(x) \lren\right]\nonumber\\
& + g(x_1) g(x_2) \left(\sum_{k \in Q} c_k\right) \left(\Delta_{g g}^{(2) C_A}(x) + \Delta_{g g}^{(2) C_F}(x)\right) 
\end{align}

\section*{Appendix B: Correction Terms}
\label{sec:apB}
\renewcommand{\theequation}{{\rm{B}}.\arabic{equation}}
\setcounter{equation}{0}

In order to present the expressions for the different corrections, we define the following distributions
\begin{align}
\mathcal{D}_i(x) =& \left[\frac{\log^i(1-x)}{1-x}\right]_+
\end{align}
which appear in the soft terms regularising the divergence of soft emission ($x\approx 1$) and defined as usual by
\begin{align}
\int_0^1 \mathcal{D}_i(x) f(x) {\rm d}x =& \int_0^1 \frac{\log^i(1-x)}{1-x} \left[f(x)-f(1)\right] {\rm d}x \, .
\end{align}

We also define an auxiliary variable to write the dependence on the factorisation scale, 
\begin{equation}\FacLog = \log\left(\frac{\mu_F^2}{Q^2}\right),\end{equation}
where $\mu_F$ is the factorisation scale and $Q$ the invariant mass of the produced $Z$.

The corrections needed for the NLO result are
\begin{align}
\Delta_{q\bar q}^{(1)} = & 8 \Dcero \FacLog+16 \Duno+\delta (x-1) (6 \FacLog+8 \zeta_2-16) \nonumber\\&
-4 \FacLog (x+1)-\frac{4 \left(x^2+1\right) \log (x)}{1-x}-8 (x+1) \log (1-x)\\
\Delta_{q g}^{(1)} = & \frac{1}{2} \left(2 \left(2 x^2-2 x+1\right) (\FacLog+2 \log (1-x)-\log (x))-7 x^2+6 x+1\right).
\end{align}

For the second NNLO, several correction terms are introduced. We denote with a $C_A$ superscript the corrections coming from the non-abelian part of the contributions (only relevant for the QCD NNLO contribution), with $n_F$ the ones that involve a sum over fermion families (relevant for the QCD and QED NNLO result), and with $C_F$ the rest of the abelian contributions.

The non-abelian contributions on the quark-antiquark channel are
\begin{align}
\Delta_{q\bar q}^{(2) C_A} =& \Delta_{q\bar q}^{\text{NS},C_A} - \Delta_{q\bar q}^{C_F-C_A/2}\\
\Delta_{q\bar q}^{(2) C_F} =& \Delta_{q\bar q}^{\text{NS},C_F} + 2 \Delta_{q\bar q}^{C_F-C_A/2}\\
\Delta_{q\bar q}^{(2) n_F} =& \frac{8}{27} \left(9 \Dcero \FacLogDos + \FacLog \left(36 \Duno -30 \Dcero\right)-36 \Dcero \zeta_2+28 \Dcero\right.\nonumber\\
&\left.-60 \Duno+36 \Ddos\right)+\frac{1}{18} \delta (x-1) \left(36 \FacLogDos-204 \FacLog-224 \zeta_2\right.\nonumber\\
&\left.+144 \zeta_3+381\right) + \frac{1}{27 (x-1)} \left(2 \left(2 \left(18 x^2 \DiLog(1-x)-(x-1)\right.\right.\right.\nonumber\\
&\left.\times \left(9 \FacLogDos (x+1)+\FacLog (6-66 x)-36 x \zeta_2+103 x-36 \zeta_2-47\right)\right)\nonumber\\
&-24 \log (1-x) \left((x-1) (3 \FacLog (x+1)-11 x+1)-6 \left(x^2+1\right) \log (x)\right)\nonumber\\
&+18 \left(4 \FacLog \left(x^2+1\right)-11 x^2+10 x-9\right) \log (x)-72 \left(x^2-1\right) \log ^2(1-x)\nonumber\\
&\left.\left.-9 \left(5 x^2+7\right) \log ^2(x)\right)\right)\footnotemark,\end{align}\footnotetext{Here we amend the result for $\Delta^{(2)}_{q\bar{q},A^2}$ given in Eq.(B.11) of ref \cite{vanNeerven:1991gh} by adding the corresponding missing $x$ factor to term $103x$ above.}

where
\begin{align}
\Delta_{q\bar q}^{\text{NS},C_A} = &  \frac{1}{180} \delta (x-1) \left(-1980 \FacLogDos-4320 \FacLog \zeta_3+11580 \FacLog\right.\nonumber\\
&\left.-432 \zeta_2^2+11840 \zeta_2+5040 \zeta_3-23025\right)-\frac{4}{27} \left(99 \Dcero \FacLogDos\right.\nonumber\\
&\left.+108 \Dcero \FacLog \zeta_2-402 \Dcero \FacLog-396 \Dcero \zeta_2-378 \Dcero \zeta_3\right.\nonumber\\
&\left.+404 \Dcero+396 \Duno \FacLog+216 \Duno \zeta_2-804 \Duno+396 \Ddos\right)\nonumber\\
&+ \frac{1}{27 (x-1)}\left[2 \left(36 \DiLog(1-x) \left(3 \FacLog \left(x^2+1\right)-7 x^2+3 x+3\right)\right.\right.\nonumber\\
&\left.\left.+99 \FacLogDos \left(x^2-1\right)+6 \FacLog (x-1) (2 x (9 \zeta_2-62)+18 \zeta_2-19)\right.\right.\nonumber\\
&\left.\left.+270 x^2 \Spence(1-x)-162 \Spence(1-x)+324 \TriLog(1-x)-450 x^2 \zeta_2-378 x^2 \zeta_3\right.\right.\nonumber\\
&\left.\left.+1139 x^2+108 x \zeta_2-1362 x+342 \zeta_2+378 \zeta_3+223\right)+12 \log (1-x) \right.\nonumber\\
&\left.\times \left(36 x^2 \DiLog(1-x)+(x-1) (66 \FacLog (x+1)+36 x \zeta_2-239 x+36 \zeta_2\right.\right.\nonumber\\
&\left.\left.-38)-6 \left(22 x^2+13\right) \log (x)\right)-18 \log (x) \left(12 \left(x^2+1\right) \DiLog(1-x)\right.\right.\nonumber\\
&\left.\left.+\FacLog \left(44 x^2+26\right)+12 x^2 \zeta_2-109 x^2+83 x+12 \zeta_2-78\right)\right.\nonumber\\
&\left.+792 \left(x^2-1\right) \log ^2(1-x)+9 \left(55 x^2+32\right) \log ^2(x)\right]
\end{align}
\begin{align}
\Delta_{q\bar q}^{\text{NS},C_F} = & \FacLog (-\Dcero (64 \zeta_2+128)+96 \Duno+192 \Ddos)+\FacLogDos (48 \Dcero\nonumber\\
&+64 \Duno)+256 \Dcero \zeta_3-\Duno (128 \zeta_2+256)+128 \Dtres+\delta (x-1) \nonumber\\
&\times \left(\FacLogDos (18-32 \zeta_2)+\FacLog (24 \zeta_2+176 \zeta_3-93)+\frac{8 \zeta_2^2}{5}\right. \nonumber\\
&\left. -70 \zeta_2-60 \zeta_3+\frac{511}{4}\right)+ \frac{1}{3 (x-1)}\left[2 \left(12 \left(\DiLog(1-x) \left(2 \FacLog \left(x^2-3\right)\right.\right.\right.\right.\nonumber\\
&\left.-4 x^2+x+3\right)+\FacLogDos \left(-\left(x^2+4 x-5\right)\right)+\FacLog (x-1) (x (4 \zeta_2+2)\nonumber\\
&+4 \zeta_2+15)+6 x^2 \Spence(1-x)+2 \Spence(1-x)-4 x^2 \TriLog(1-x)+6 \TriLog(1-x)-8 x^2 \zeta_2\nonumber\\
&\left.-16 x^2 \zeta_3+6 x^2+16 x \zeta_2-15 x-8 \zeta_2+16 \zeta_3+9\right)+6 \log (1-x) \nonumber\\
&\times\left(6 \left(x^2-3\right) \DiLog(1-x)-(x-1) \left(8 \FacLogDos (x+1)-4 \FacLog (x-7)\right.\right.\nonumber\\
&\left.-x (16 \zeta_2+3)-16 (\zeta_2+4)\right)+4 \left(\FacLog \left(9 x^2+5\right)-7 x^2+11 x-4\right) \log (x)\nonumber\\
&\left.-12 \left(2 x^2+1\right) \log ^2(x)\right)+12 \log (x) \left(3 \left(x^2+1\right) \DiLog(1-x)+\FacLogDos \left(3 x^2+1\right)\right.\nonumber\\
&\left.+\FacLog \left(-3 x^2+10 x-1\right)-12 x^2 \zeta_2+6 x^2-19 x-4 \zeta_2-1\right)\nonumber\\
&-6 \log ^2(1-x) \left(8 (x-1) (3 \FacLog (x+1)-2 x+2)-\left(39 x^2+23\right) \log (x)\right)\nonumber\\
&-6 \left(\FacLog \left(9 x^2+3\right)-6 x^2+8 x-2\right) \log ^2(x)-96 \left(x^2-1\right) \log ^3(1-x)\nonumber\\
&\left.\left.+\left(25 x^2+11\right) \log ^3(x)\right)\right]
\end{align}

\begin{align}
\Delta_{q\bar q}^{C_F-C_A/2} = & \frac{1}{3 (x-1)}\left(12 \DiLog(1-x) \left(2 \FacLog x^2+2 \FacLog-9 x^3+4 x^2\right.\right.\nonumber\\
&\left.+4 \left(x^2+1\right) \log (1-x)+2 \left(2 x^3+6 x^2-4 x-1\right) \log (x)+9 x-1\right)\nonumber\\
&+72 (x-1) (x+1)^2 \DiLog(-x) (\log (x)-2 \log (x+1)+1)+84 \FacLog x^2\nonumber\\
&+12 \FacLog x^2 \log ^2(x)-24 \FacLog x^2 \log (x)-180 \FacLog x\nonumber\\
&+12 \FacLog \log ^2(x)+60 \FacLog \log (x)+96 \FacLog+96 x^3 \Spence(1-x)\nonumber\\
&-144 x^3 \Spence(-x)+300 x^2 \Spence(1-x)-144 x^2 \Spence(-x)-192 x \Spence(1-x)\nonumber\\
&+144 x \Spence(-x)+12 \Spence(1-x)+144 \Spence(-x)-24 x^3 \TriLog(-x)-72 x^2 \TriLog(1-x)\nonumber\\
&-24 x^2 \TriLog(-x)+24 x \TriLog(-x)-24 \TriLog(1-x)+24 \TriLog(-x)+36 x^3 \zeta_2\nonumber\\
&+24 x^3 \zeta_2 \log (x)-72 x^3 \zeta_2 \log (x+1)-78 x^3+4 x^3 \log ^3(x)-90 x^3 \log ^2(x)\nonumber\\
&-72 x^3 \log (x) \log ^2(x+1)+60 x^3 \log ^2(x) \log (x+1)+72 x^3 \log (x) \log (x+1)\nonumber\\
&+36 x^2 \zeta_2+24 x^2 \zeta_2 \log (x)-72 x^2 \zeta_2 \log (x+1)-240 x^2+6 x^2 \log ^3(x)\nonumber\\
&+24 x^2 \log (1-x) \log ^2(x)+54 x^2 \log ^2(x)-72 x^2 \log (x) \log ^2(x+1)\nonumber\\
&+60 x^2 \log ^2(x) \log (x+1)+168 x^2 \log (1-x)-48 x^2 \log (1-x) \log (x)\nonumber\\
&+42 x^2 \log (x)+72 x^2 \log (x) \log (x+1)-36 x \zeta_2-24 x \zeta_2 \log (x)+72 x \zeta_2 \log (x+1)\nonumber\\
&-24 \zeta_2 \log (x)+72 \zeta_2 \log (x+1)+762 x-12 x \log ^3(x)-14 \log ^3(x)+84 x \log ^2(x)\nonumber\\
&+72 x \log (x) \log ^2(x+1)-60 x \log ^2(x) \log (x+1)+24 \log (1-x) \log ^2(x)\nonumber\\
&-93 \log ^2(x)+72 \log (x) \log ^2(x+1)-60 \log ^2(x) \log (x+1)-360 x \log (1-x)\nonumber\\
&+162 x \log (x)-72 x \log (x) \log (x+1)+192 \log (1-x)+120 \log (1-x) \log (x)\nonumber\\
&\left.-276 \log (x)-72 \log (x) \log (x+1)-36 \zeta_2-444\right)
\end{align}

The singlet contributions for the (anti)quark-(anti)quark channel include terms arising from identical initial quarks, and non-identical ones. These are given by the following expressions.

\begin{align}
\Delta_{qq}^{(2) \text{id}} = & 2 \Bigg(\FacLog \left(-\frac{1}{x+1}\left(4 \left(x^2+1\right) \left(4 \DiLog(-x)-\log ^2(x)+4 \log (x+1) \log (x)+2 \zeta_2\right)\right)\right.\nonumber\\
&\left.+8 (x+1) \log (x)-16 (x-1)\right. \bigg)-\frac{1}{3 (x+1)}\left(4 \left(x^2+1\right) \left(-18 \DiLog(1-x) \log (x)\right.\right.\nonumber\\
&+12 \DiLog(-x) (2 \log (1-x)-2 \log (x)+\log (x+1))-24 \TriLog\left(\frac{1-x}{x+1}\right)+24 \TriLog\left(\frac{x-1}{x+1}\right)\nonumber\\
&-24 \Spence(1-x)+12 \Spence(-x)+24 \TriLog(1-x)+6 \TriLog(-x)-9 \zeta_2 \log (x)+12 \zeta_2 \log (1-x)\nonumber\\
&+6 \zeta_2 \log (x+1)+2 \log ^3(x)-6 \log (1-x) \log ^2(x)-21 \log (x+1) \log ^2(x)\nonumber\\
&\left.\left.+6 \log ^2(x+1) \log (x)+24 \log (1-x) \log (x+1) \log (x)+3 \zeta_3\right)\right)+(1-x) \nonumber\\
&\times\Big(-16 \DiLog(-x) \log (x+1)-16 \Spence(-x)+8 \TriLog(-x)+4 \zeta_2 \log (x)-8 \zeta_2 \log (x+1)\nonumber\\
&-\frac{2}{3} \log ^3(x)+4 \log (x+1) \log ^2(x)-8 \log ^2(x+1) \log (x)+32 \log (1-x)+8 \zeta_3-34\Big)\nonumber\\
&+4 (x+1) (2 \DiLog(-x)+4 \log (1-x) \log (x)+2 \log (x) \log (x+1)+\zeta_2)\nonumber\\
&+8 (x+3) \DiLog(1-x)-4 (3 x+1) \log ^2(x)+2 (7 x-9) \log (x) \Bigg) -\frac{4}{3} (x-1)^2 \nonumber\\
&\times\left(6 \DiLog(1-x) (2 \log (x)+3)+12 \Spence(1-x)-12 \TriLog(1-x)+2 \log ^3(x)+9 \log ^2(x)\right) \nonumber\\
&+4 (6 x-7) \log (x)-26 x^2+56 x-30\\
\Delta_{qq}^{(2) \text{non-id}} = & \frac{1}{54 x}\left(36 \DiLog(1-x) (x (12 \FacLog (x+1)+x (8 x+15)+39)+6 x (x+1) \right.\nonumber\\
&\times(4 \log (1-x)+\log (x))+16)+6 x \log (x) \left(18 \FacLogDos (x+1)+36 \log (1-x) \right.\nonumber\\
&\times\left(2 (\FacLog+3) x+2 \FacLog+2 (x+1) \log (1-x)+4 x^2+3\right)\nonumber\\
&\left.+18 \FacLog \left(4 x^2+6 x+3\right)+20 x^2-72 (x+1) \zeta_2-48 x+345\right)\nonumber\\
&+(x-1) \left(-18 \FacLogDos (x (4 x+7)+4)+24 \log (1-x) \right.\nonumber\\
&\times(-3 \FacLog (x (4 x+7)+4)-3 (x (4 x+7)+4) \log (1-x)+(17-22 x) x-22)\nonumber\\
&\left.-12 \FacLog (x (22 x-17)+22)+703 x^2+72 (x (4 x+7)+4) \zeta_2-1895 x-116\right)\nonumber\\
&-9 x \log ^2(x) (24 \FacLog (x+1)+48 (x+1) \log (1-x)+5 (x (8 x+15)+3))\nonumber\\
&\left.+432 x (x+1) (3 \Spence(1-x)-2 \TriLog(1-x))+162 x (x+1) \log ^3(x)\right)
\end{align}
It is worth noticing the sign difference in the non-identical vectorial contribution for $qq$ initial state, with respect to $q\bar q$.
\begin{align}
\Delta_{q \bar q}^{(2)\text{non-id, V}} = & -\Delta_{q q}^{\text{non-id, V}} = \frac{1}{3 x}\left(2 \left(12 \DiLog(-x) \left(2 \left(2 x^2+5\right) \log (x)+5 x (x+1)-6 (x (x+2)+2)\right.\right.\right.\nonumber\\
&\left.\times \log (x+1)\right)-6 x \DiLog(1-x) (4 x+(x-10) \log (x)-5)+6 (3 x ((x-2) \TriLog(1-x)\nonumber\\
&-4 (x+2) \Spence(-x))+4 (x (x+2)+2) \Spence(1-x)+2 (10-3 x) x \TriLog(-x))\nonumber\\
&-144 \Spence(-x)+24 \TriLog(1-x)-120 \TriLog(-x)+30 x (x (\zeta_2+4)+\zeta_2-4)\nonumber\\
&+6 \log (x+1) (5 \log (x) (2 x (x+1)+(x (x+2)+2) \log (x))-6 (x (x+2)+2) \zeta_2)\nonumber\\
&-x \log (x) (-6 x \zeta_2+48 x+x \log (x) (4 \log (x)+39)-60 \zeta_2+60)-18 ((x-6) x\nonumber\\
&\left.\left.+4) \zeta_3-36 (x (x+2)+2) \log (x) \log ^2(x+1)\right)\right)\\
\Delta_{q \bar q}^{(2) \text{non-id, A}} = & 4 \DiLog(1-x) ((3 x+2) \log (x)+1)+8 \DiLog(-x) (x+8 \log (x)-6 (x+2) \log (x+1)\nonumber\\
&+1)+4 (4 (x+2) \Spence(1-x)-x (12 \Spence(-x)+\TriLog(1-x)-10 \TriLog(-x)))\nonumber\\
&-96 \Spence(-x)+\frac{2}{3} \left(-72 \TriLog(-x)+6 (x (\zeta_2+9 \zeta_3+4)+\zeta_2-6 \zeta_3-4)\right.\nonumber\\
&+6 \log (x+1) \left(-6 (x+2) \zeta_2+5 (x+2) \log ^2(x)+2 (x+1) \log (x)\right)\nonumber\\
&+\log (x) (6 (5 x \zeta_2+2 \zeta_2-2)-x \log (x) (4 \log (x)+3))\nonumber\\
&\left.-36 (x+2) \log (x) \log ^2(x+1)\right)+8 \TriLog(1-x)
\end{align}

The correction terms that appear in the $qg$ and $q\gamma$ channels are given by the following expressions.
\begin{align}
\Delta_{q g}^{(2) C_A} = & \frac{1}{2} \left(\frac{2}{9} \FacLog \Big(36 \left(2 x^2+6 x+3\right) \DiLog(1-x)-36 \left(2 x^2+2 x+1\right) (\DiLog(-x)\right.\nonumber\\
&+\log (x) \log (x+1))-72 \left(2 x^2-x+1\right) \zeta_2+73 x^2+54 \left(2 x^2-2 x+1\right) \log ^2(1-x)\nonumber\\
&+6 \left(-71 x^2+54 x+\frac{8}{x}+9\right) \log (1-x)+18 \left(28 x^2-2 x+3\right) \log (x)\nonumber\\
&+36 \left(-2 x^2+10 x+1\right) \log (1-x) \log (x)-12 x+\frac{44}{x}-36 (3 x+1) \log ^2(x)-87\Big)\nonumber\\
&-4 \left(2 x^2+2 x+1\right) \left(4 \DiLog(-x) (\log (1-x)-\log (x))+2 \TriLog(-x)-4 \TriLog\left(\frac{1-x}{x+1}\right)\right.\nonumber\\
&\left.+4 \TriLog\left(\frac{x-1}{x+1}\right)-3 \log (x+1) \log ^2(x)+4 \log (1-x) \log (x+1) \log (x)\right)\nonumber\\
&+\frac{4}{3} \left(44 x^2+90 x+\frac{16}{x}+33\right) \DiLog(1-x)+8 \left(5 x^2+10 x+7\right) \DiLog(1-x) \log (1-x)\nonumber\\
&+8 \left(4 x^2+5 x+1\right) (\DiLog(-x)+\log (x) \log (x+1))+8 x \DiLog(1-x)\nonumber\\
&+8 x (7-2 x)\DiLog(1-x) \log (x)+\frac{2}{3} \FacLogDos \Big(-31 x^2+6 \left(2 x^2-2 x+1\right) \log (1-x)\nonumber\\
&+24 x+\frac{4}{x}+6 (4 x+1) \log (x)+3\Big)+8 \left(4 x^2+16 x+9\right) \Spence(1-x)-4 \left(12 x^2+34 x\right.\nonumber\\
&\left.+15\right) \TriLog(1-x)+\frac{4}{3} \left(107 x^2-84 x-\frac{8}{x}+15\right) \zeta_2-32 \left(2 x^2-x+1\right) \zeta_2 \log (1-x)\nonumber\\
&-4 \left(2 x^2+4 x+1\right) \zeta_3+\frac{1837 x^2}{27}+\frac{26}{3} \left(2 x^2-2 x+1\right) \log ^3(1-x)\nonumber\\
&+\frac{4}{3} \left(-77 x^2+63 x+\frac{8}{x}+6\right) \log ^2(1-x)+4 \left(-6 x^2+22 x+1\right) \log (x) \log ^2(1-x)\nonumber\\
&+4 \left(2 x^2-14 x-3\right) \log ^2(x) \log (1-x)-\left(\frac{346 x^2}{3}+5\right) \log ^2(x)+\frac{2}{9} \Big(74 x^2+75 x\nonumber\\
&+\frac{88}{x}-210\Big) \log (1-x)+20 \left(13 x^2-2 x+1\right) \log (x) \log (1-x)-\frac{2}{9} \left(457 x^2+12 x\right.\nonumber\\
&\left.-354\right) \log (x)+16 x (2 x-5) \zeta_2 \log (x)-\frac{1226 x}{9}+\frac{116}{27 x}+\frac{2}{3} (20 x+9) \log ^3(x)\nonumber\\
&\left.-4 x \log ^2(x)+8 x \log (x) \log (1-x)-8 x \log (x)+\frac{539}{9}\right)\\
\Delta_{q g}^{(2)C_F} = & \frac{1}{2} \Bigg(\FacLog \left(-48 x^2 \DiLog(1-x)+\left(2 x^2-2 x+1\right) \left(36 \log ^2(1-x)-8 \zeta_2\right)+22 x^2\right.\nonumber\\
&+8 \left(4 x^2-2 x+1\right) \log ^2(x)-4 \left(23 x^2-34 x+8\right) \log (1-x)+2 \left(46 x^2-40 x+5\right)\nonumber\\
&\left.\times \log (x)-8 \left(16 x^2-10 x+5\right) \log (1-x) \log (x)-68 x+24\right)+\left(2 x^2-2 x+1\right) \nonumber\\
&\left(-16 \DiLog(-x) \log (x)+32 \TriLog(-x)-16 \zeta_2 \log (1-x)+\frac{70}{3} \log ^3(1-x)+100 \zeta_3\right)\nonumber\\
&+2 \left(40 x^2-28 x+3\right) \DiLog(1-x)-4 \left(26 x^2-6 x+3\right) \DiLog(1-x) \log (1-x)\nonumber\\
&-16 \left(3 x^2+2 x-1\right) (\DiLog(-x)+\log (x) \log (x+1))+8 (x-3) \DiLog(1-x)+4 (1-2 x)\nonumber\\
&\times \DiLog(1-x) \log (x)+3 \FacLogDos \left(\left(8 x^2-8 x+4\right) \log (1-x)+\left(-8 x^2+4 x-2\right)\right.\nonumber\\
&\left.\times \log (x)+4 x-1\right)-4 \left(34 x^2-22 x+11\right) \Spence(1-x)+4 \left(18 x^2+2 x-1\right) \TriLog(1-x)\nonumber\\
&+4 \left(-12 x^2+2 x+5\right) \zeta_2+24 \left(4 x^2-2 x+1\right) \zeta_2 \log (x)-\frac{305 x^2}{2}-\frac{1}{3} \left(52 x^2-34 x\right.\nonumber\\
&\left.+17\right) \log ^3(x)-\frac{1}{2} \left(4 x^2-68 x+35\right) \log ^2(x)+8 \left(10 x^2-6 x+3\right) \log (1-x) \log ^2(x)\nonumber\\
&-6 \left(22 x^2-14 x+7\right) \log ^2(1-x) \log (x)-2 \left(63 x^2-80 x+23\right) \log ^2(1-x)-\left(174 x^2\right.\nonumber\\
&\left.-245 x+59\right) \log (x)+4 \left(48 x^2-50 x+13\right) \log (1-x) \log (x)+2 \left(88 x^2-147 x+38\right)\nonumber\\
&\times \log (1-x)-12 (x-1)+233 x-4 (x-3) \log ^2(x)+(28-44 x) \log (x)+8 (x-3)\nonumber\\
&\times \log (1-x) \log (x)+24 (x-1) \log (1-x)-\frac{181}{2}\Bigg)
\end{align}

The last correction terms correspond to the ones contributing to the $gg$, $g\gamma$ and $\gamma \gamma$ channels.
\begin{align}
\Delta_{g g}^{(2) C_A} = & \frac{C_A^2}{C_A^2-1} \, \frac{1}{3} \big(-8 (x+1)^2 \DiLog(-x) (9 \log (x)-6 \log (x+1)-2)-24 (x-1)^2 \Spence(1-x)\nonumber\\
&+48 (x+1)^2 \Spence(-x)+72 x (x+2) \TriLog(-x)+72 \TriLog(-x)+x (x (8 \zeta_2+48 \zeta_3+191)\nonumber\\
&+16 (\zeta_2+6 \zeta_3-9))+24 \zeta_2 \log (x+1)+4 \log (x+1) \left(6 x (x+2) \zeta_2+(x+1)^2 (4\right.\nonumber\\
&\left.-9 \log (x)) \log (x)\right)+24 (x+1)^2 \log (x) \log ^2(x+1)+2 \log (x) (-x (75 x+38)\nonumber\\
&+(x (25 x+2)-2) \log (x)-6)+8 \zeta_2+48 \zeta_3-47\big)\\
\Delta_{q g}^{(2) C_F} = &\FacLog \left((2 x+1)^2 (2 \log (x) (\log (x)-4 \log (1-x))-8 \DiLog(1-x))-67 x^2+60 \right.\nonumber\\
&\left.x+16 (x-1) (3 x+1) \log (1-x)+2 (1-4 (x-2) x) \log (x)+7\right)-4 (x+1)^2\nonumber\\
&\times \left(\log (x+1) \left(4 \DiLog(-x)-3 \log ^2(x)+2 \log (x+1) \log (x)+2 \zeta_2\right)+4 \Spence(-x)\right)\nonumber\\
&-4 (2 x+1)^2 (\DiLog(1-x) (4 \log (1-x)+\log (x))-4 \TriLog(1-x)+\log (1-x) \nonumber\\
&\times(2 \log (1-x)-\log (x)) \log (x))+4 (2 x (7 x-2)-5) \DiLog(1-x)+8 (x (x+4)+2)\nonumber\\
&\times \DiLog(-x) \log (x)+8 (x+1) (\DiLog(-x)+\log (x) \log (x+1))-2 \FacLogDos \nonumber\\
&\times\left(-6 x^2+4 x+(2 x+1)^2 \log (x)+2\right)-8 (x (7 x+10)+1) \Spence(1-x)+8 ((x-2) x\nonumber\\
&-1) \TriLog(-x)+98 x^2+4 (3 (3-4 x) x+5) \zeta_2+4 (10 x (x+1)+3) \zeta_2 \log (x)+4 (2 (x-1) x\nonumber\\
&-1) \zeta_3-66 x-\frac{2}{3} (8 x (x+1)+3) \log ^3(x)-2 (x+1) (4 x+3) \log ^2(x)+16 (x-1) \nonumber\\
&\times (3 x+1) \log ^2(1-x)+(x (105 x-64)-23) \log (x)+4 (1-4 (x-2) x) \log (1-x) \nonumber\\
&\times\log (x)-2 (x-1) (67 x+7) \log (1-x)-32
\end{align}
\end{appendices}
\end{document}